\documentclass[fleqn,usenatbib]{mnras}

\usepackage{newtxtext,newtxmath}

\usepackage[T1]{fontenc}

\DeclareRobustCommand{\VAN}[3]{#2}
\let\VANthebibliography\thebibliography
\def\thebibliography{\DeclareRobustCommand{\VAN}[3]{##3}\VANthebibliography}

\usepackage{graphicx}	
\usepackage{amsmath}	
\usepackage{standalone}
\usepackage{longtable}
\usepackage{afterpage}
\usepackage{tabularx}
\usepackage{multirow}
\usepackage{xcolor}
\usepackage{ulem}

\newcommand{\src}{V1674~Her}

\newcommand{\swift}{\textit{Swift}}
\newcommand{\astr}{\textit{AstroSat}}
\newcommand{\nic}{\textit{NICER}}
\newcommand{\chan}{\textit{Chandra}}

\title[AstroSat view of \src]{Soft X-ray and FUV observations of Nova Her 2021 (V1674~Her) with \astr}

\author[Bhargava et al.]
{Yash Bhargava,$^{1,2}$\thanks{E-mail: yash.bhargava\_003@tifr.res.in}
Gulab Chand Dewangan,$^{1}$
G. C. Anupama,$^{3}$
U. S. Kamath,$^{3}$
L. S. Sonith,$^{3}$\newauthor
Kulinder Pal Singh,$^{4}$
J. J. Drake,$^{5,6}$
A. Beardmore,$^{7}$
G. J. M. Luna, $^{8}$
M. Orio, $^{9,10}$
K. L. Page $^{7}$
\\
$^{1}$Inter-University Centre for Astronomy and Astrophysics, Ganeshkhind, Pune 411007, India\\
$^{2}$Department of Astronomy and Astrophysics, Tata Institute of Fundamental Research, 1 Homi Bhabha Road, Colaba, Mumbai 400005, India\\
$^{3}$Indian Institute of Astrophysics, II Block Koramangala, Bengaluru 560034, India\\
$^{4}$Department of Physical Sciences, Indian Institute of Science Education and Research Mohali, Knowledge City, Sector 81, SAS Nagar, Punjab 140306, India\\
$^{5}$Center for Astrophysics$\,|\,$Harvard and Smithsonian, 60 Garden Street, Cambridge, MA 02138, USA\\
$^{6}$Lockheed Martin, 3251 Hanover St, Palo Alto, CA 94304, USA\\
$^{7}$School of Physics and Astronomy, The University of Leicester, University Road, Leicester LE1 7RH, UK\\
$^{8}$CONICET/Universidad Nacional de Hurlingham, Av. Gdor. Vergara 2222, Villa Tesei, Buenos Aires, Argentina\\
$^{9}$INAF-Osservatorio di Padova, Vicolo Osservatorio 5, I-35122 Padova, Italy\\
$^{10}$Department of Astronomy, University of Wisconsin, 475 N, Charter Street, Madison, WI 53706, USA
}

\date{Accepted XXX. Received YYY; in original form ZZZ}

\pubyear{2023}

\begin{document}\label{firstpage}
\pagerange{\pageref{firstpage}--\pageref{lastpage}}
\maketitle

\begin{abstract}%
Nova Her 2021 or V1674 Her was one of the fastest novae to be observed so far. We report here the results from our timing and spectral studies of the source observed at multiple epochs with \astr. We report the detection of a periodicity in the source in soft X-rays at a period of 501.4--501.5~s which was detected with high significance after the peak of the super-soft phase, but was not detected in the far ultraviolet (FUV) band of \astr. The shape of the phase-folded X-ray light curves has varied significantly as the nova evolved. The phase-resolved spectral studies reveal the likely presence of various absorption features in the soft X-ray band of 0.5--2~keV, and suggest that the optical depth of these absorption features may be marginally dependent on the pulse phase. Strong emission lines from Si, N and O are detected in the FUV, and their strength declined continuously as the nova evolved and went through a bright X-ray state.  
\end{abstract}%




\section{Introduction}

Novae are interacting binaries with a white dwarf accreting hydrogen rich material from a (predominantly) low mass, late-type main sequence star. This material forms an accretion disc around the white dwarf. 
With the accretion of sufficient material a thermonuclear runaway (TNR) occurs in the matter that is electron degenerate \citep{starrfield2012BASI...40..419S, Starrfield2020ApJ...895...70S}, or under ideal gas conditions for accretion rates $\ga 3 \times 10^{-9}$~M$_\odot$~yr$^{-1}$ \citep{Chomiuk2021ARA&A..59..391C}.
The TNR causes the nova eruption, an event that is accompanied by the ejection of the accreted material with velocities ranging from a few hundreds to several thousands km s$^{-1}$ and an increase in the luminosity of the object by $\sim 5-19$ magnitudes \citep{vogt1990ApJ...356..609V, ozd2018MNRAS.476.4162O, kawash2021ApJ...910..120K}. 
Nova outbursts typically occur for accretion rates of $\dot M\le10^{-8}\,\rm{M}_\odot\,\rm{yr}^{-1}$,
while in some systems, where the white dwarf mass is higher, nova outbursts can occur for a higher accretion rate. Systems with one recorded outburst are termed classical novae, while those with multiple outbursts are termed recurrent novae. The white dwarf in some nova systems could be magnetic. A strong magnetic field prevents the formation of an accretion disc, with accretion happening along the magnetic poles of the white dwarf (polars), while in some cases the strength of the magnetic field is weaker, enough to just disrupt and truncate the inner regions of the accretion disc (intermediate polars), with accretion in such cases happening via accretion curtains on the magnetic poles of the white dwarf \citep[see][for a recent review]{Della2020A&ARv..28....3D}.

The classical nova \src\ (Nova Her 2021; TCP J18573095+1653396) was discovered on 2021 June 12.537~UT by Seiji Ueda at a visual magnitude of 8.4, and on the rise \citep[\url{http://www.cbat.eps.harvard.edu/unconf/followups/J18573095+1653396.html,}][]{li2021ATel14705....1L}. 
Data from the All-Sky Automated Survey for Supernovae (ASAS-SN) revealed a pre-discovery detection, on 2021 June 12.1903  \citep{Aydi2021ATel14710....1A}. Spectroscopic observations on 2021 June 12.84 by \citet{munari2021ATel14704....1M} showed the presence of broad P Cygni absorption features of hydrogen Balmer lines at a velocity $\sim 3000$ km s$^{-1}$ confirming it to be a nova. Based on magnitudes in the AAVSO database, \citet{woodward2021ApJ...922L..10W} estimate the maximum occurred on June 12.96 at $V=6.14$ mag. With the estimated decline rates of $t_2=1.2$~day, $t_3=2.2$~days and $t_6=14$~days corresponding to change in magnitudes $\Delta m=2,3$ and $6$ relative to the maximum, respectively \citep{quimby2021RNAAS...5..160Q, woodward2021ApJ...922L..10W}, nova \src\ is the fastest Galactic nova known to date. 

Early UV and optical spectra obtained during the first two-three days since discovery showed the presence of Fe II lines \citep{Aydi2021ATel14710....1A, albanese2021ATel14718....1A, kuin2021ATel14736....1K, woodward2021ApJ...922L..10W}, that were not present in the later spectra obtained $\ga 5$ days since discovery \citep{kuin2021ATel14736....1K, wang2021ATel14737....1W, woodward2021ApJ...922L..10W}.
Early phase spectroscopic observations also indicated changes in the line profiles on timescales of a day \citep{Aydi2021ATel14710....1A}. In addition to the absorption system noted by \citet{munari2021ATel14704....1M}, faster components developed, with P Cygni absorptions at blueshifted velocities $> 5000$~km s$^{-1}$. These absorption systems are likely due to multiple outflows \citep{Aydi2021ATel14710....1A}. Near-Infrared (NIR) spectra indicated the emergence of coronal lines as early as $+11$~d, the earliest observed for any classical nova \citep{woodward2021ApJ...922L..10W}. Strong neon lines were present in the late time ($\sim$month) spectra \citep{wagner2021ATel14746....1W, ochner2021ATel14805....1O} suggesting \src\ to be an ONe nova, i.e. a nova occurring on an ONe WD.

An uncatalogued $\gamma$-ray source at the nova position was detected in the Fermi-LAT data of 0.1--300~GeV obtained during  2021 June 12.0 to 12.96 \citep{li2021ATel14705....1L, li2021ATel14707....1L} with $F(0.1-300 GeV)= 1.1\pm 0.3 \times 10^{-6}$ photons cm$^{-2}$ s$^{-1}$ ($4.4 \pm 1.3 \times 10^{-10}$ erg cm$^{-2}$ s$^{-1}$), and photon index of $2.6 \pm0.3$. \src\ was also detected at radio frequencies, as early as $\sim3$ days since discovery \citep{sokolovsky2021ATel14731....1S}. A detailed multi-wavelength study from $\gamma-$rays to radio wavelengths by \citet{sokolovsky2023MNRAS.521.5453S} indicated the presence of internal shocks that were responsible for the $\gamma-$ray emission and the early X-ray emission in the \textit{NuSTAR} (3--30 keV) and \textit{Swift} (0.3--10 keV) observations. In the radio, while the early emission corresponded to thermal, it later developed to synchrotron emission due to the shocks. \src\ entered the supersoft X-ray (SSS) phase,  caused by nuclear burning of the residual material on the WD surface, at $\sim 19$ days \citep{page2021ATel14747....1P}, and lasted until $\sim 60$ days.

\citet{mroz2021ATel14720....1M} reported a period of 501.4277s in the $r-$band (pre-outburst) light curve based on observations with the Zwicky Transient Facility (ZTF) since 2018, and attributed it to the spin of the WD. \citet{patterson2022ApJ...940L..56P} reported the detection of a 501.486(5)s period in the post-outburst optical light curve, beyond $\sim 12$ days after discovery. No periodicity was seen in the near-IR by \citet{hansen2021RNAAS...5..244H} in their observations during days 3--6  after discovery. \citet{maccarone2021ATel14776....1M} detected X-ray oscillations with a 
period of 503.9 s and an amplitude of 0.6 to 1.4 times the
mean count rate based on {\it Chandra} High-Resolution Camera photometry obtained on day 28 of the outburst. \citet{Drake2021ApJ...922L..42D} reported a period of $501.72\pm 0.11$~s in the X-ray based on {\it Chandra} Low Energy Transmission Grating Spectrometer observations on day 37. 

The post-outburst increase in the spin period has been attributed to the sudden loss of high-angular-momentum gas from the rotating, magnetic white dwarf due to the nova outburst \citep{Drake2021ApJ...922L..42D, patterson2022ApJ...940L..56P}. \textit{Swift} X-ray observations obtained $\sim 10$ months after outburst indicated the X-ray light curve continued to be modulated at this period \citep{Page2022ATel15317....1P}. \citet{patterson2022ApJ...940L..56P} also reported the detection of a 0.152921(3) day period in the optical light curve, attributed to the orbital period of the system.  A similar period is also seen in the \textit{TESS} lightcurve with an additional periodicity at 0.537~d of an unknown origin \citep{Luna2023arXiv231002220L}.  \citet{lin2022MNRAS.517L..97L} report the detection of this period (0.153~d) in the X-ray bands based on \nic\ observations obtained about a month since discovery.

We present here results based on \astr\ observations of \src\ in the soft X-ray and Far-UV (FUV) bands during its SSS phase. We describe the observations and the data reduction methods in Section~\ref{sec:obs}. The timing and spectral analysis of the observations are detailed in Sections~\ref{sec:timing} and \ref{sec:spec} respectively. The results and conclusions are reported in Section~\ref{sec:results}. 

\section{Observations and Data Reduction}\label{sec:obs}
\astr\ \citep{Singh2014SPIE.9144E..1SS} is the first Indian multi-wavelength astronomy mission operated by the Indian Space Research Organisation (ISRO). It was launched on 2015 September 28 into a low Earth orbit. \astr\ carries four co-aligned instruments --- the Ultraviolet Imaging Telescope \citep[UVIT;][]{Tandon2017AJ....154..128T, Tandon2020AJ....159..158T}, the Soft X-ray Telescope \citep[SXT;][]{Singh2016SPIE.9905E..1ES, Singh2017JApA...38...29S}, the Large Area Proportional Counters \citep[LAXPC;][]{yadav2016SPIE.9905E..1DY, antia2021JApA...42...32A} and the Cadmium Zinc Telluride Imager \citep[CZTI;][]{vadawale2016SPIE.9905E..1GV}. 

\astr\  observed \src\ on four epochs. Due to the supersoft nature of the source, only SXT and UVIT were  used for the observation. The details of the observations and various modes/filters used are reported in Table~\ref{tab:obs_log_astr}. 
In the article, the observations are referred to by the days elapsed since the start of the optical outburst  (2021 June 12.1903 or equivalently MJD 59377.1903) of the source (also indicated in Table~\ref{tab:obs_log_astr}). 
\begin{table*}
    \centering
     \caption{Log of \astr\ observations. Throughout the article, the observations are referred to by the days elapsed since the detection of the source (also mentioned in the parentheses in the first column). The combined exposure of UVIT filter/gratings for individual observation is less than the corresponding SXT exposure as UVIT has stricter observational constraints than SXT.}\label{tab:obs_log_astr}
    \begin{tabular}{|l|ll|cc|r|r|r|r|}\hline
         OBSID & Instrument & Mode/Filter & Start time & End time & Exposure  & Count Rate & Flux \\ 
          & & & MJD & MJD & in ks & in cts/s & in  ergs cm$^{-2}$ s$^{-1}$        \\\hline
         \multirow{4}{*}{T04\_019T01\_9000004516 (Day 24)} & SXT & PC &59401.636736 & 59402.067998 & 12.57 & 1.199 & 8.4 $\times$10$^{-11}$\\
          & UVIT & BaF2 & 59401.641166 & 59401.652870 & 1 & 7.79 & \\
          & UVIT &  FUV-G1 & 59401.654610 & 59401.852731 & 5.05 & 6.97 & \\
          & UVIT &  FUV-G2 & 59401.854462 & 59402.250974 & 4 & 6.52 & \\ \hline
         \multirow{4}{*}{T04\_026T01\_9000004556 (Day 37)} & SXT & PC & 59414.355441 & 59415.057187 & 20.49 & 10.27  & 6.6 $\times$10$^{-10}$\\ 
         & UVIT & BaF2 & 59414.359530 & 59414.564107 & 5 & 2.98 &  \\
          & UVIT &  FUV-G1 & 59414.565846 & 59414.917615 & 9.95 & 2.69 & \\
          & UVIT &  FUV-G2 & 59414.976885 & 59414.987880 & 0.95 & 2.21 & \\ \hline
         \multirow{2}{*}{ T04\_026T01\_9000004560 (Day 38)} & SXT & PC & 59415.437903 & 59415.728501 & 10.66 & 11.07 & 7.22 $\times$10$^{-10}$\\ 
          & UVIT &  FUV-G2 & 59415.439995 & 59415.723665 & 8.7 & 2.05 & \\ \hline
         \multirow{3}{*}{ T04\_026T01\_9000004624 (Day 54)} & SXT & PC & 59431.269271 & 59431.621536 & 11.84 & 5.751 & 3.85 $\times$10$^{-10}$\\ 
          & UVIT & BaF2 & 59431.339212 & 59431.426182 & 4.45 & 0.976 & \\
          & UVIT &  FUV-G1 & 59431.475899 & 59431.620301 & 3.55 & 1.32 & \\\hline
         
    \end{tabular}
   
\end{table*}

\subsection{SXT}

\begin{figure}
    \centering
    \includegraphics[width=\columnwidth]{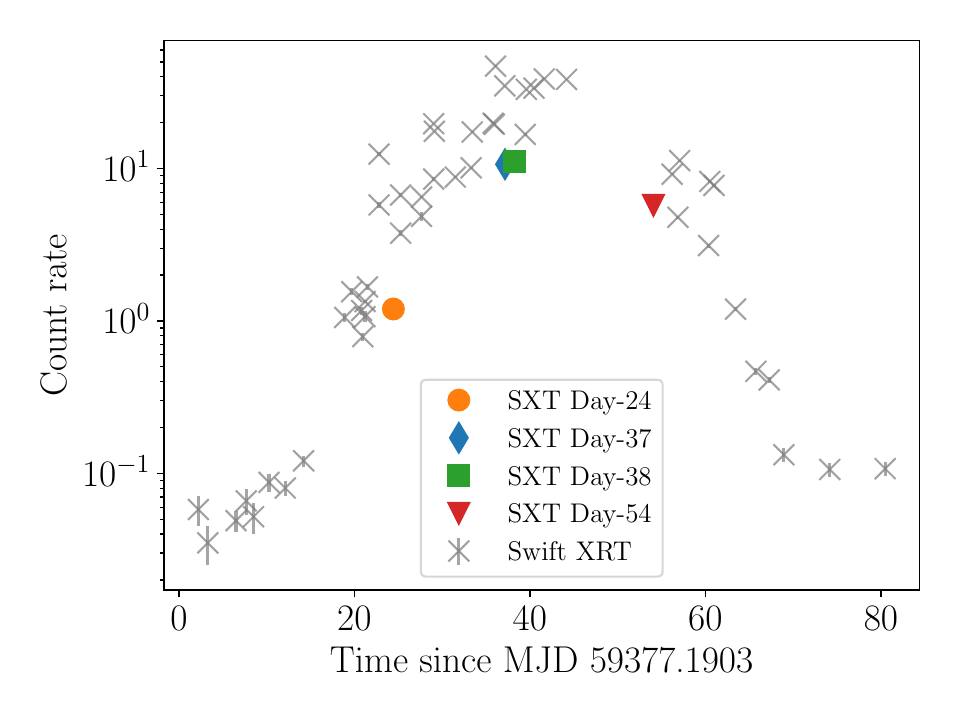}
    \caption{Evolution of the \src\ in soft X-rays as seen by \swift-XRT shown as grey crosses. The epochs of SXT observations are shown in different colours and symbols. The UVIT observations were simultaneous with SXT observations as indicated in Table~\ref{tab:obs_log_astr}. }\label{fig:sxt_xrt_evolution}
\end{figure}

\begin{figure*}
    \centering
    \includegraphics[width=2\columnwidth]{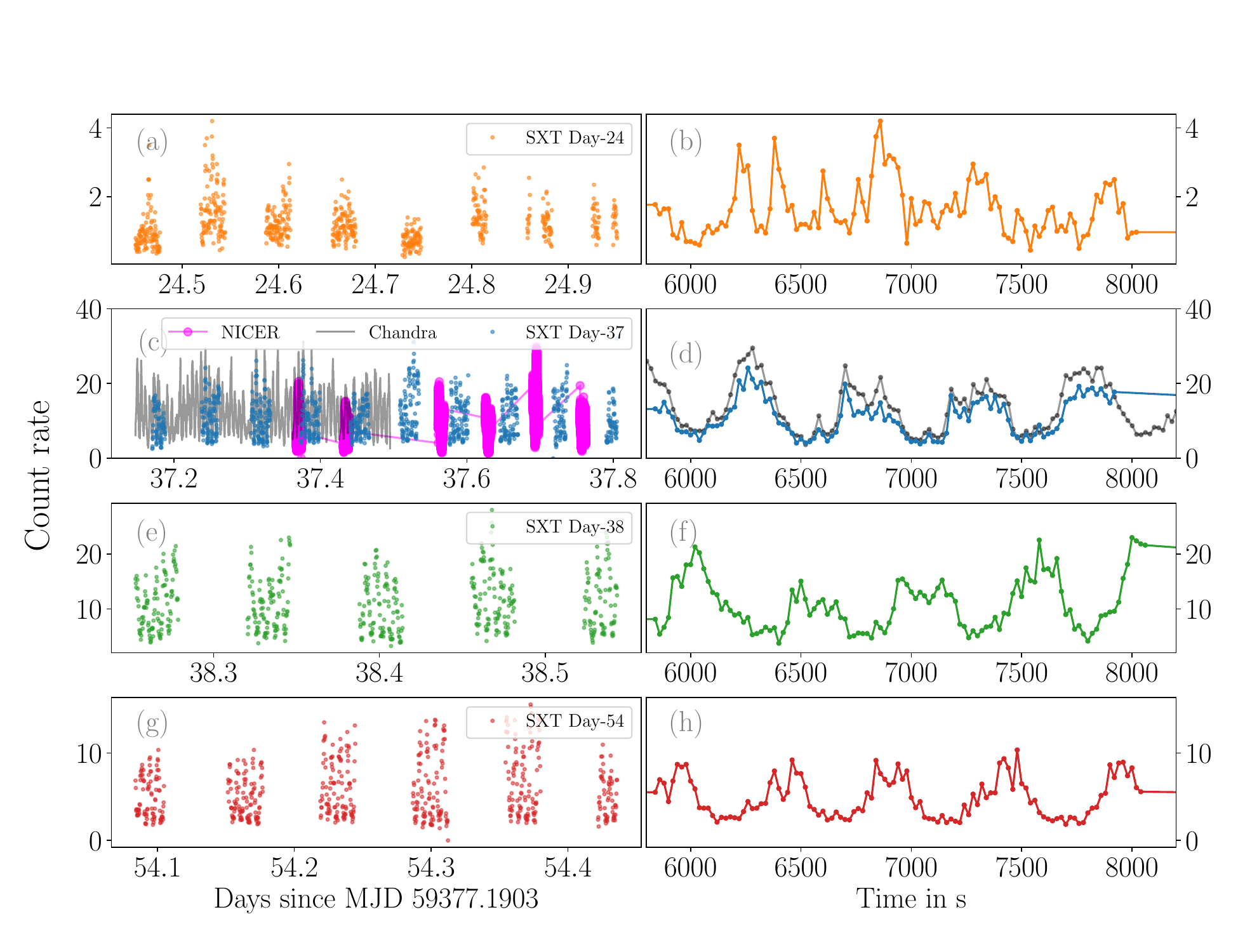}
    \caption{Soft X-ray lightcurves of different epochs of observation of \src. The lightcurves are extracted in the 0.35--2 keV energy range for SXT observations binned at 20 s. \nic\ observed the source contemporaneously with SXT observation on Day 37 (see table~\ref{tab:obs_log_astr}) and the corresponding lightcurve is shown in magenta. The NICER count rates have been divided by 100 to compare the variations in the different observations. \chan\ also observed the source \citep[][]{Drake2021ApJ...922L..42D}  with a significant overlap with SXT observation which is shown as the grey solid line. Right panels [(b), (d), (f), (h)] show the lightcurves at different epochs for a single \astr\ orbit (the second orbit in the left panels). Observations on days 37, 38 and 54 after source detection show clear periodic variation with differences evident across consecutive pulses.   }\label{fig:sxt_lc_nicer}
\end{figure*}

SXT \citep{Singh2016SPIE.9905E..1ES, Singh2017JApA...38...29S} observed the source on 4 epochs as seen in Figure~\ref{fig:sxt_xrt_evolution} with individual lightcurves shown in Figure~\ref{fig:sxt_lc_nicer}. All the observations were conducted in the Photon Counting (PC) mode.  The minimum time resolution of SXT in this mode is 2.3775~s \citep{Singh2016SPIE.9905E..1ES, Singh2017JApA...38...29S}.  The SXT observations were reduced to level 2 format using the SXTPIPELINE v1.4b as provided by SXT Payload operation centre (POC). The pipeline also includes required calibration files. The data from individual \astr\ orbits were merged accounting for the overlap and repeated events using a \textsc{julia} based tool provided by SXT-POC. Light curves and spectra were extracted from the merged events file using \texttt{xselect} which is a part of the HEASARC suite. We used a circular region with a radius of 16$^\prime$ centred on the  source image accounting for the typical large point-spread function  of the instrument. The count rate of the source is typically less than the threshold for pile-up, and therefore we do not exclude the central region for our studies. Due to the large point-spread function of the instrument, simultaneous background estimation is not possible and thus a standard background (provided by POC) is used. The source is fainter than background level beyond 2 keV, thus the spectral analysis is limited to 0.5--2 keV. The standard response files provided by POC are used and the ancillary response files are modified for the selected region using the code provided by the POC. For the determination of the pulse period, we use lightcurves with shortest possible time resolution (i.e. 2.3775~s).  
The latter part of Day 24 observation was contaminated with a solar flare and therefore the corresponding interval was excluded from the analysis.

\subsection{UVIT}

\begin{figure*}
    \centering
    \includegraphics[width=2\columnwidth]{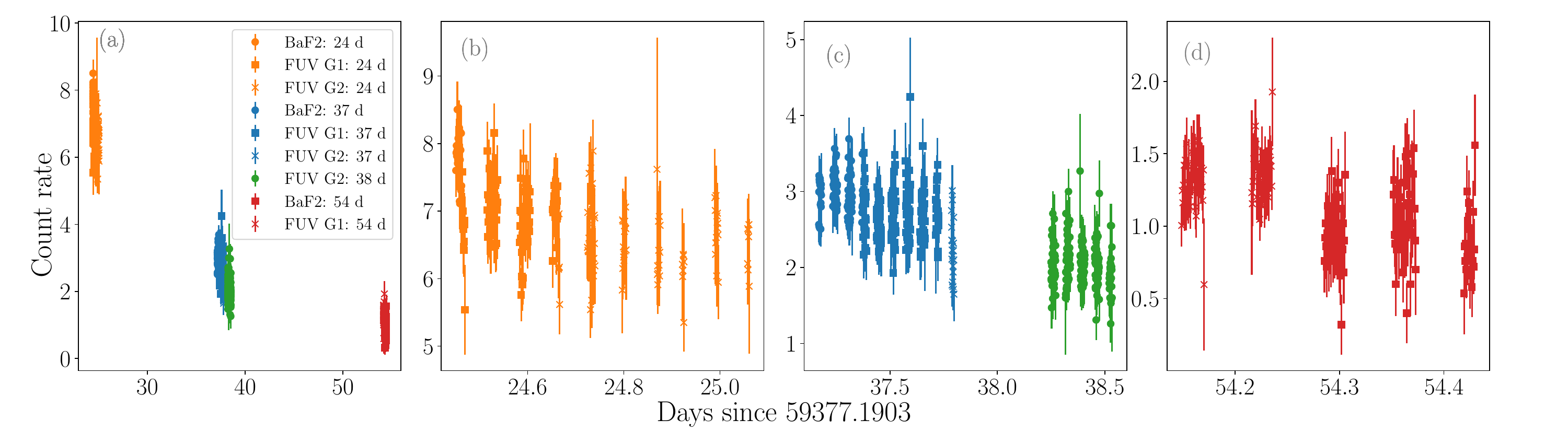}
    \caption{UV lightcurves as seen by \astr-UVIT. The lightcurves are binned at 50~s. Panel a shows the evolution of the source in UV wavelengths as the outburst progresses while panels b, c and d show a close-up of the lightcurves at each of the three epochs. }\label{fig:uvit_lc}
\end{figure*}

The UVIT \citep{Tandon2017AJ....154..128T, Tandon2020AJ....159..158T} consists of twin telescopes, one of which  offers far ultraviolet sensitivity and is known as the FUV channel (1200--1800\AA). The second telescope constitutes two channels providing sensitivity in the near ultraviolet (NUV channel; 2000--3000\AA) and visible (VIS channel; 3200--5500\AA) bands. The FUV and NUV channels are equipped with a number of broadband filters and slit-less gratings, and operate in the photon counting mode.  The FUV/NUV channels are capable of high-resolution imaging (FWHM $\sim 1-1.5{\rm~arcsec}$) and timing capability with an accuracy of few millisecond per 1000~s. Here we use the UVIT data on \src\ acquired with the FUV channel using the BaF2 filter (F154W, $\lambda_{mean} = 1541$~\AA, $\Delta\lambda=380$~\AA)  and the two gratings FUV-Grating1 (FUV-G1) and FUV-Grating2 (FUV-G2). The two slit-less gratings are mounted in the FUV filter wheel so that their dispersion axes are nearly orthogonal.
We obtained the level1 data from the \astr\ archive\footnote{\url{https://astrobrowse.issdc.gov.in/astro_archive/archive/Home.jsp}}, and processed them using the CCDLAB pipeline \citep{postma2017PASP..129k5002P}. 
 We corrected the orbit-wise cleaned event lists for pointing drifts, and aligned them for each observation. 
We converted these event lists to event files compatible with the HEASARC tool xselect\footnote{\url{https://heasarc.gsfc.nasa.gov/ftools/xselect/}}, and merged them to obtain a single events file for the entire exposure in the BaF2 filter.  We then used xselect to extract light curves with 50~s bins from the merged event files using a circular region of radius 10$^{\prime\prime}$ centred at the source position. We also extracted background light curves from the source-free regions, and corrected the source light curves for background contributions.
The lightcurves of the observations are shown in Figure~\ref{fig:uvit_lc}. 

We also aligned the orbit-wise images and merged them into a single image for each observation.  For the spectral extraction, we used the UVITTools.jl package\footnote{\url{https://github.com/gulabd/UVITTools.jl}}, and followed the procedures and tools described in \citet{2021JApA...42...49D} and \citet{kumar2023ApJ...950...90K}. We used the  FUV grating order $-2$ as it is well calibrated due to its maximum intensity.
We first located the zeroth order position of the source in the grating images, we then used the centroids along the spatial direction at each pixel to find the dispersion direction for the $-2$ order.
We used a 50-pixel width along the cross-dispersion direction and extracted the one-dimensional count spectra for the FUV gratings in the $-2$ order. Following a similar procedure we also extracted background  count spectra from source-free regions, and corrected the source spectra for the background contribution. We list the background-subtracted net source count rates for FUV gratings in the $-2$ order in Table~1.  We then performed  wavelength and flux calibration and produced the fluxed spectra.

\section{Periodicity analysis}\label{sec:timing}

\begin{figure}
    \centering
    \includegraphics[width=\columnwidth]{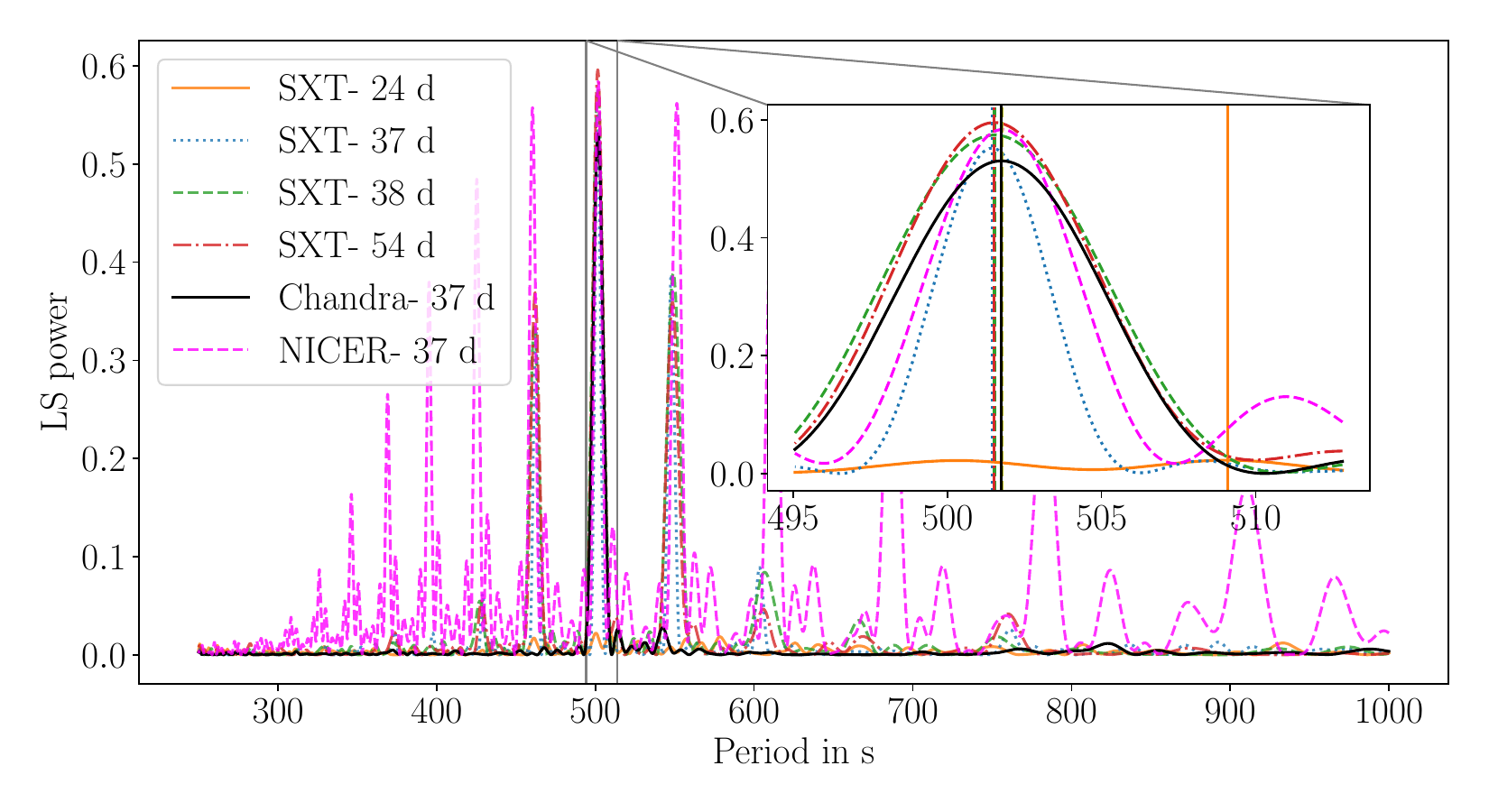}
    \caption{LS periodogram of \src\ in soft X-rays. The colour scheme of the periodograms is kept similar to Figure~\ref{fig:sxt_xrt_evolution} and the legend indicates the number of days passed since the source detection. 
    The periodogram for \nic\ observation which was contemporaneous with observation on day 37 is shown in magenta. The inset shows the periodogram close to 501~s with vertical lines highlighting the peak positions.  }\label{fig:sxt_lsp}
\end{figure}

The X-ray lightcurves of  Days 37, 38 and 54  show a clear signature of periodic behaviour. The  Day 24 lightcurve  (which is right at the beginning of the SSS emission), although variable, does not show periodic changes in the count rate. The period suggested by \citet{patterson2021ATel14856....1P} and \citet{Drake2021ApJ...922L..42D} is $\sim$500~s which is only a factor of 4 smaller than a continuous interval probed by SXT observations. Therefore we use Lomb-Scargle (LS) techniques\footnote{as implemented in \textsc{astropy}} \citep{lomb1976Ap&SS..39..447L, scargle1982ApJ...263..835S,vanderplas2012cidu.conf...47V,vanderplas2015ApJ...812...18V, vdp2018ApJS..236...16V} to determine the period. To search for the periodicity around the suggested period, we computed the LS power in a limited frequency range (0.001--0.004~Hz, with 100000 linearly spaced bins) allowing the code to pick the frequency with the highest power. We find that in observations on Days 37, 38 and 54 the periodogram  detects significant power close to 0.002 Hz (i.e. $\sim$500~s) while in observation on Day 24 no clear signal is detected at similar frequency. The periodograms for SXT observations are shown in Figure~\ref{fig:sxt_lsp}. The $x$-axis is converted to period (measured in seconds) for clarity. The additional peaks detected in the periodogram are alias peaks arising due to windowing and data gaps present in the \astr\ observations.

The observation on day 37 was contemporaneous with \chan\ and \nic\ observations. The overlap of the observations is shown in Figure~\ref{fig:sxt_lc_nicer}. The \nic\ observation (OBSID: 4202260107) typically covered some of the data gaps in the \astr\ observation. To compare the period detection in SXT, \chan\ and \nic, we computed similar periodograms using \chan\ and \nic\ observations which are also shown in Figure~\ref{fig:sxt_lsp}.  We report the period observed in the SXT observations in Table~\ref{tab:period_x-ray}. We determine the statistical uncertainty in the period by using the method described in Appendix~\ref{app:error}.

\begin{table}
\centering
\caption{Detected period from the LS periodogram from \astr-SXT observations.  The errorbars reported correspond to the standard deviation of the periods measured from the simulated lightcurves as described in Appendix~\ref{app:error}}\label{tab:period_x-ray}
    \begin{tabular}{l|r} \hline
    Observation day & Period \\ \hline
    Day 37 &  501.44 $\pm0.01$\\
    Day 38 &  501.53 $\pm0.05$ \\
    Day 54 &  501.52 $\pm0.04$ \\ \hline
\end{tabular}
\end{table}

\begin{figure}
    \centering
    \includegraphics[width=\columnwidth]{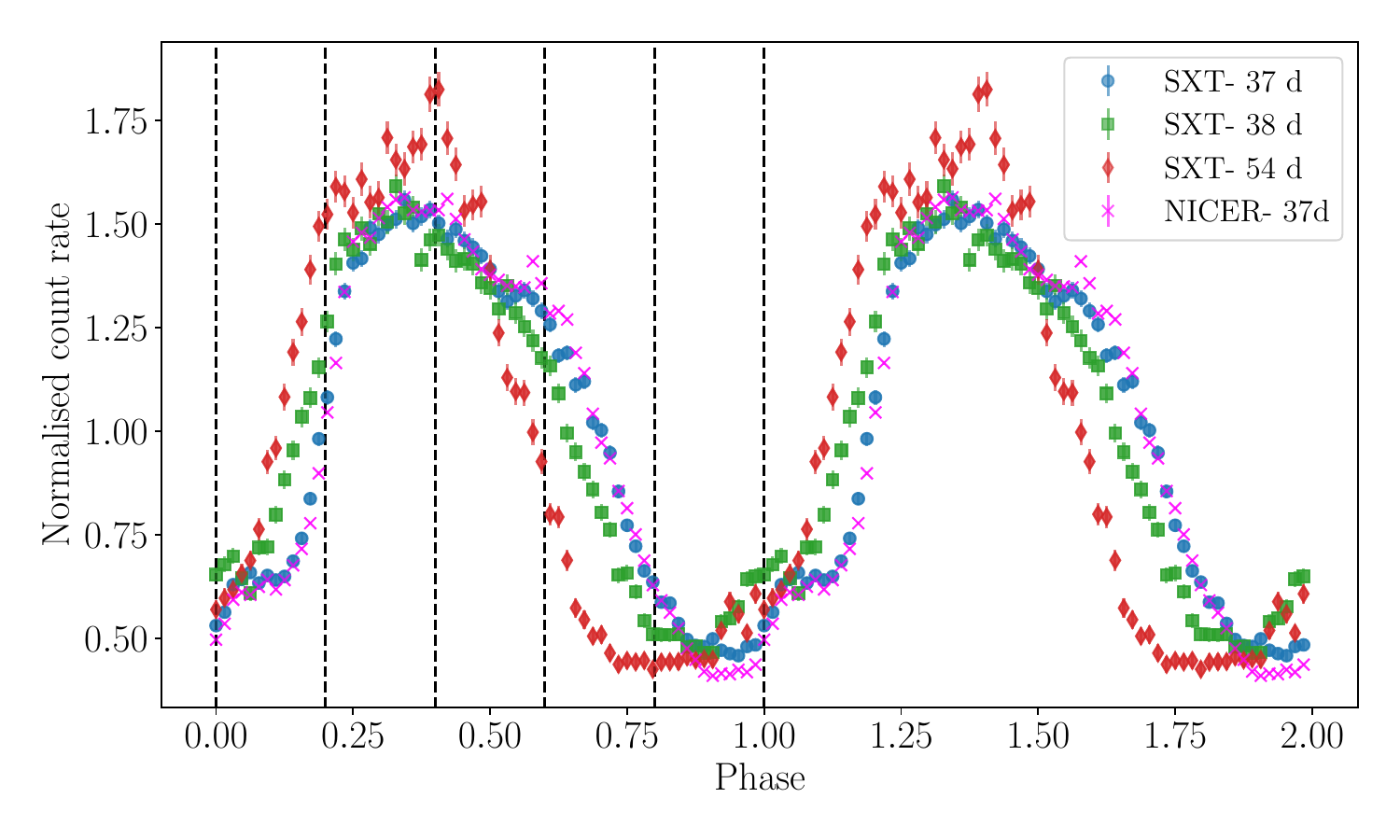}
    \caption{Phase-folded lightcurves for different SXT observations and NICER observation which is contemporaneous with observation on day 37. The lightcurves were folded with phase 0.0 corresponding to MJD 59414.00115 and the 501.7~s period suggested by \citet{Drake2021ApJ...922L..42D}. The lightcurves are normalised to indicate the pulse profile and two intervals of the period are shown for clarity. The intervals for the pulse-resolved spectral analysis are shown as dashed lines in the first pulse. }\label{fig:phase_folded_sxt}
\end{figure}

The phase-folded lightcurves for different observations are shown in Figure~\ref{fig:phase_folded_sxt}. The lightcurves are folded at the period suggested by \citet{Drake2021ApJ...922L..42D} and with a common epoch of MJD 59414.00115. The phase-folded lightcurve of contemporaneous NICER observation is also plotted for comparison.

The measured pulse periods in different SXT observations and in contemporaneous \chan\ and \nic\ observations are statistically different. The period seen in \src\ is $\sim$501~s which is only a factor of 4 less than the typical continuous exposure in SXT while in \nic\ observations \citep[analysed here and in][]{Orio2022ApJ...932...45O} it is of similar duration. Due to variations in the individual pulses, the accurate determination of the period may also depend on the duration of the continuous exposure and an accurate determination of the time of arrival of individual pulses. We tested the effect of various sampling windows on period determination using the LS technique. We sampled the \chan\ lightcurve with a window function corresponding to typical data gaps and continuous exposures as seen in SXT and \nic\ observations. The period determined using the LS technique on the sampled lightcurve is also significantly different\footnote{The statistical uncertainty on the period is typically 0.01--0.05~s while the differences in the period is 0.3~s.} from the period determined from the complete lightcurve. The periodogram constructed from the \chan\ observation sampled in the exact simultaneous interval as SXT-37 d observation shows a period identical to the SXT observation.

\begin{figure}
    \centering
    \includegraphics[width=\columnwidth]{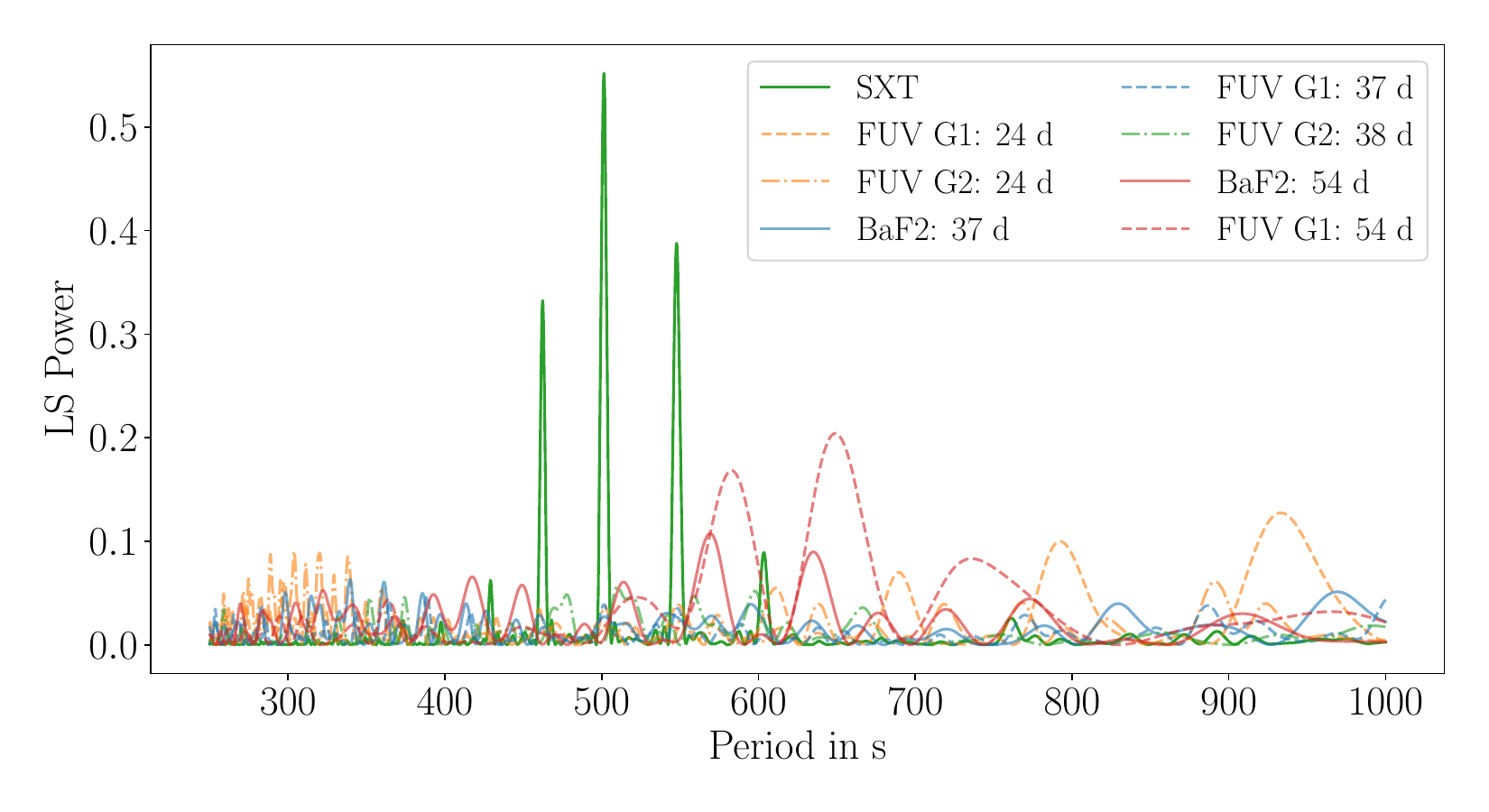}
    \caption{LS periodograms computed for individual UV observations. The colour scheme of the Figure is kept consistent with Figure~\ref{fig:uvit_lc}. The LS periodogram from SXT-38 d is also shown for reference. }\label{fig:uvit_lsp}
\end{figure}

The UV lightcurves do not indicate any periodicity close to 500~s. We quantify this by computing the LS periodogram in a similar frequency range as in Figure~\ref{fig:sxt_lsp}. The periodograms computed for the UV observations are shown in Figure~\ref{fig:uvit_lsp}.
The periodograms do not show a significant peak close to 501~s and the highest peak observed in the current frequency range for any UV observation has a false alarm probability $\gtrsim$50\% (in comparison the false alarm rate in case of SXT observations is $\lesssim$10$^{-4}$)\footnote{The false alarm probabilities in the paper are estimated for the same range of frequencies in which the periodogram is computed using the approach suggested by \citet{Baluev2008MNRAS.385.1279B} and implemented in \textsc{astropy}}. And UV observations at nearby times (e.g. BaF2 or  FUV-G1 on Day 54) do not show similar peaks in their periodograms. Thus, we conclude that the power seen in the UVIT LS periodograms is not intrinsic to the source but arises due to Poisson variation. 

To place upper limits on the detectable pulse fraction \citep[peak to trough pulse fraction = $\lbrack$maximum flux $-$ minimum flux$\rbrack$/$\lbrack$maximum flux + minimum flux$\rbrack$ as defined in][]{dhillon2009MNRAS.394L.112D} in current UV  observations, we simulated 500 lightcurves with similar noise statistics as the current observations and injected sinusoidal counts of varying pulse fractions. The noise realisations were computed by sampling the lightcurve for each time bin from a Poisson distribution with a mean defined by the counts in that bin. LS periodograms were computed from the lightcurve corresponding to each pulse fraction and the noise realisation in the frequency range of 0.001--0.004 Hz and false alarm probability was determined for the highest peak in each periodogram (irrespective of the difference between the period of the highest peak and the injected period). Assuming a conservative limit of false alarm probability of 0.01\footnote{ In the absence of a periodic signal in the lightcurve, the observed LS power is expected 1\% of the time}, for a detection of injected pulsations, we determine a corresponding upper limit of 0.05 on the observable pulse fraction.   Additionally, we estimate the LS power in an interval around the suggested optical period \citep[0.153~d; ][]{patterson2022ApJ...940L..56P} spanning a decade in the frequency and for 10000 bins in the UV lightcurves to check if the UV lightcurves indicate any orbital modulation. We do not see any strong peak in the LS periodogram and the highest peak in that period interval is consistent with a false alarm probability of 0.99. Notably, the LS power around the orbital period is much lower than the typical LS power in Figure~\ref{fig:uvit_lsp} and thus a conservative upper limit on the orbital modulation is 0.05 (peak to trough pulse fraction).

\section{Spectral analysis}\label{sec:spec}

\subsection{X-ray spectral analysis}
\begin{figure}
    \centering
    \includegraphics[width=\columnwidth]{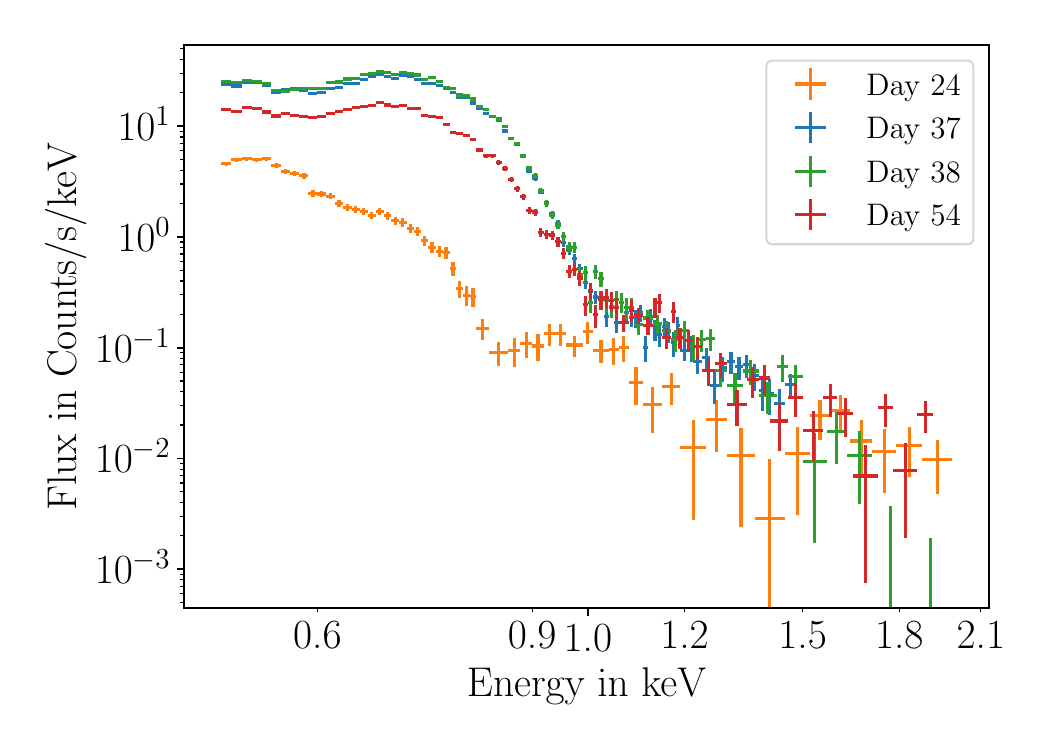}
    \caption{Pulse averaged X-ray spectrum for each epoch. The colour scheme has been kept identical to Figure~\ref{fig:sxt_xrt_evolution}}\label{fig:sxt_spec_evol}
\end{figure}

Spectral analysis was carried out with XSPEC version 12.12.1
\citep{xspec} distributed with the HEASOFT package (version 6.30.1). 
We model the time-averaged spectrum to test the evolution of the source in X-rays. The time-averaged spectra at all the epochs are depicted in Figure~\ref{fig:sxt_spec_evol}.  Following the \textit{Swift}-XRT analysis of \citet{Drake2021ApJ...922L..42D}, we fit the spectra  in the 0.5--2~keV range with a combination of a thermal blackbody (\texttt{bbody}) and multiple absorption edges (\texttt{edge}). In case of the SXT observation on day 24.5, the flux was low and thus the spectrum beyond 1~keV was dominated by background. We therefore restrict our modelling to 0.5--1~keV.   We also applied a neutral H absorption (\texttt{tbabs}) with a fixed column density of 2.9$\times10^{21}$~cm$^{-2}$ \citep{Drake2021ApJ...922L..42D}  with abundances from \citet{aspl2009ARA&A..47..481A} and cross sections from \citet{vern1996ApJ...465..487V}. The parameters of the fits are reported in Table~\ref{tab:xray_spec_pars}. 
We find that the absorption edges are at 0.61~keV (N VII), 0.77~keV (O VII), 0.87~keV (O VIII), 0.96~keV (F VIII), 1.14~keV (F IX/Ne IX) and 1.38~keV (Ne X). The edge identification is done based on the reports from \citet{Drake2021ApJ...922L..42D} and the Atomic Database\footnote{AtomDB, \url{http://www.atomdb.org/Webguide/webguide.php?z1=0&z0=7}} We note that not all edges are required to describe the fainter spectra due to lack of statistics.

\begin{figure}
    \centering
    \includegraphics[width=\columnwidth]{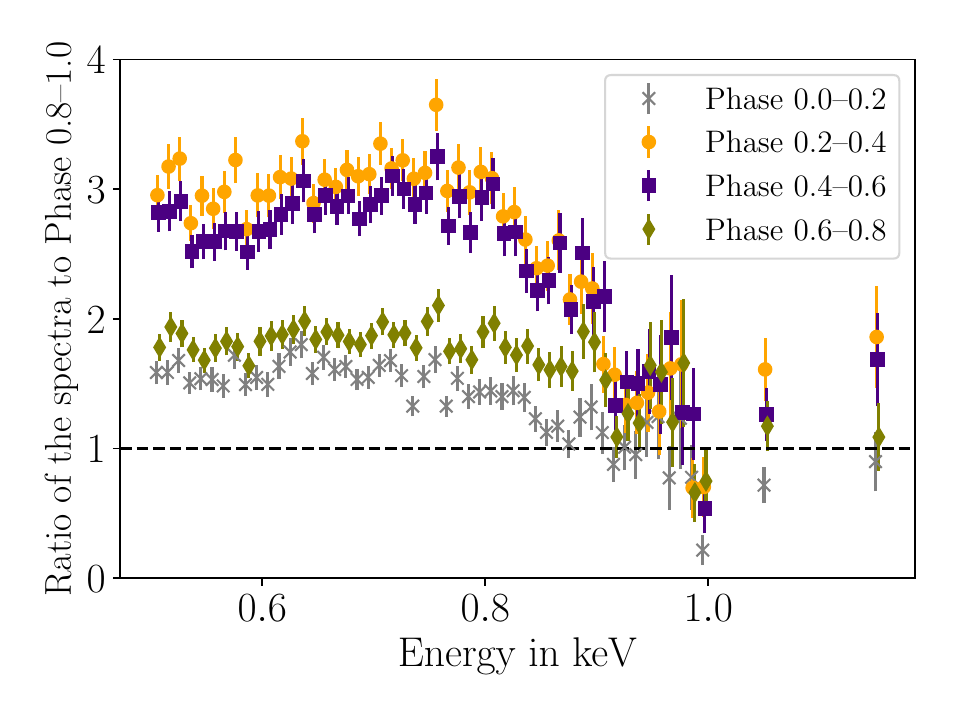}
    \caption{The ratio of the pulse phase resolved spectra with that of the pulse phase (0.8--1.0) to highlight the differences in pulse phase spectra in a model independent manner. The pulse-phase resolved spectra for this figure are extracted from the observation on day 37 after the detection of the source. }\label{fig:ratio_spec}
\end{figure}

\begin{figure}
    \centering
    \includegraphics[width=\columnwidth]{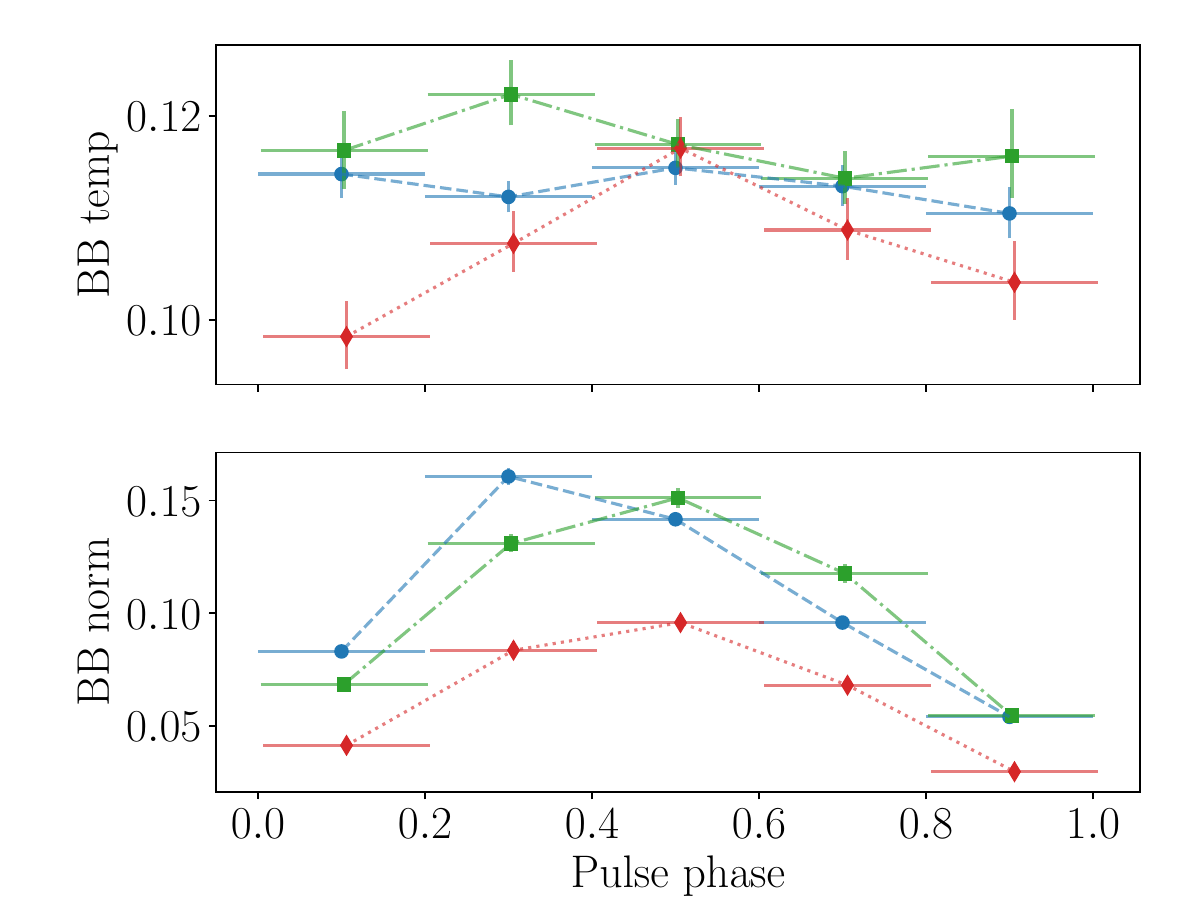}
    \caption{Evolution of the blackbody component as a function of the pulse phase. The points have been shifted in phase slightly to show the extent of 1$\sigma$ confidence intervals correctly. The colour scheme denoting the different observations is kept identical to that of Figure~\ref{fig:sxt_xrt_evolution}. }\label{fig:bb_evol}
\end{figure}

\begin{figure}
    \centering
    \includegraphics[width=\columnwidth]{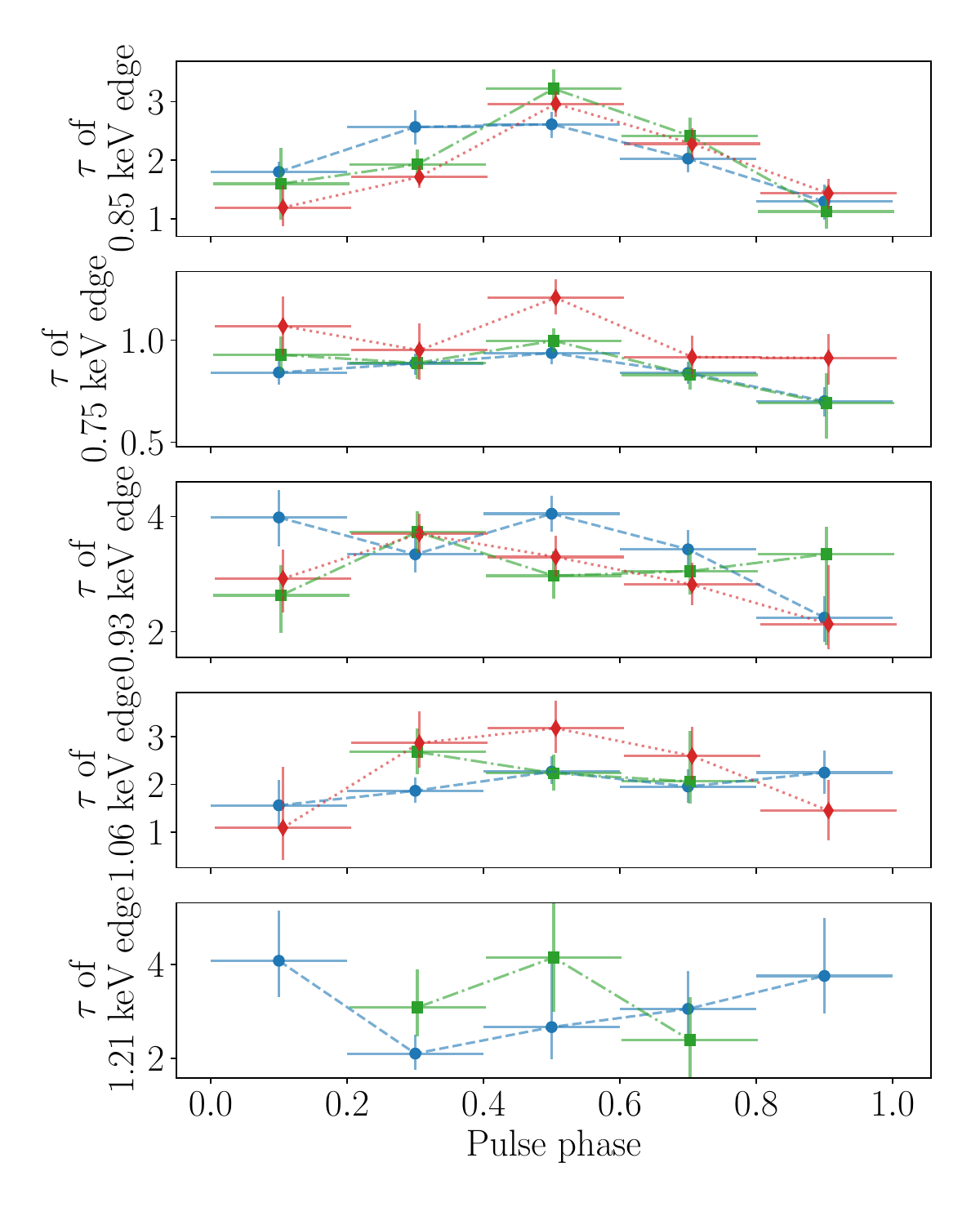}
    \caption{Evolution of the optical depth of the various edges seen in the source as a function of the pulse phase. The points have been shifted in phase slightly to show the extent of 1$\sigma$ confidence intervals correctly. The colour scheme is kept identical to that of Figure~\ref{fig:sxt_xrt_evolution}. }\label{fig:opt_depth_evol}
\end{figure}

The clear pulse evolution motivated us to investigate the variation of the spectral properties as a function of the pulse profile. We divided the pulse profile into five equal phase bins  (see Figure~\ref{fig:phase_folded_sxt} for the bin edges) and extracted the corresponding GTIs for observations on day 37, 38 and 54. The corresponding phase-resolved spectra were extracted using \texttt{xselect}. To highlight the differences in the spectra of individual pulse phases, we plot the ratio of the pulse phase resolved spectra to the phase 0.8--1.0 for the observation on day 37 in the Figure~\ref{fig:ratio_spec}.
The pulse-phase resolved spectra were modelled similar to the time-averaged spectra and the parameters of the fits are reported in Table~\ref{tab:xray_spec_pars}.  
We note that the minor fluctuations in the edge energies or the blackbody temperature ($\approx$20~eV) may not be real as they can be caused by variations in the bias voltage of detector. We depict the pulse phase dependence of the blackbody component in Figure~\ref{fig:bb_evol} and of the optical depth of various edges in Figure~\ref{fig:opt_depth_evol}. 

\begin{table*}
    \centering
    \caption{Spectral parameters for time-averaged and pulse phase resolved analysis  at different epochs of the outburst. }\label{tab:xray_spec_pars}
    \begin{tabular}{llc|c|ccccc}
    \hline
       Component  & Parameter & Unit & Time averaged &\multicolumn{5}{c}{Pulse phase}  \\ 
         & & & & 0--0.2 & 0.2--0.4 & 0.4--0.6 & 0.6--0.8 & 0.8--1 \\ \hline
         \multicolumn{9}{c}{Day 24} \\ \hline
        \multirow{2}{*}{Blackbody} & kT & keV & $0.118_{-0.008}^{+0.009}$ & --	&	--	&	--&	-- &	-- \\
          & Norm& & $0.022_{-0.001}^{+0.001} $ & --	&	--	&	--&	-- &	--  \\
         \multirow{2}{*}{Edge} & E & keV & $0.611_{-0.002}^{+0.002}$	& --	&	--	&	--&	-- &	--  \\
         & $\tau$ &&						$1.8_{-0.2}^{+0.2}$ & --	&	--	&	--&	-- &	--  \\
         \multirow{2}{*}{Edge} & E & keV & $0.722_{-0.007}^{+0.005}$	& --	&	--	&	--&	-- &	--  \\
         & $\tau$ &&						$1.9_{-0.4}^{+0.4}$ & --	&	--	&	--&	-- &	--  \\
         \multirow{2}{*}{Edge} & E & keV & $0.793_{-0.008}^{+0.008}$	& --	&	--	&	--&	-- &	--  \\
         & $\tau$ &&						$4.6_{-0.4}^{+0.4}$ & --	&	--	&	--&	-- &	--  \\
         & $\chi^2$/dof & & 56.5/31 & --	&	--	&	--&	-- &	--  \\
         \hline
         \multicolumn{9}{c}{Day 37} \\ \hline
        \multirow{2}{*}{Blackbody} & kT & keV & $0.111_{-0.001}^{+0.001}$ & $0.114_{-0.002}^{+0.003}$	&	$0.112_{-0.002}^{+0.002}$	&	$0.115_{-0.002}^{+0.002}$	&	$0.113_{-0.002}^{+0.002}$	&	$0.110_{-0.002}^{+0.003}$ \\
          & Norm& & $0.103_{-0.002}^{+0.002} $ & $0.083_{-0.002}^{+0.003}$	&	$0.161_{-0.004}^{+0.004}$	&	$0.142_{-0.003}^{+0.003}$	&	$0.096_{-0.003}^{+0.003}$	&	$0.054_{-0.002}^{+0.002}$ \\
         \multirow{2}{*}{Edge} & E & keV & $0.770_{-0.003}^{+0.002}$	& $0.749_{-0.004}^{+0.004}$	&	$0.773_{-0.003}^{+0.003}$	&	$0.772_{-0.004}^{+0.003}$	&	$0.764_{-0.004}^{+0.004}$	&	$0.758_{-0.006}^{+0.006}$ \\
         & $\tau$ &&						$0.87_{-0.04}^{+0.03}$ & $0.84_{-0.06}^{+0.06}$	&	$0.89_{-0.05}^{+0.05}$	&	$0.94_{-0.05}^{+0.05}$	&	$0.84_{-0.05}^{+0.05}$	&	$0.70_{-0.07}^{+0.07}$ \\
         \multirow{2}{*}{Edge} & E & keV & $0.869_{-0.003}^{+0.003}$	& $0.848_{-0.005}^{+0.005}$	&	$0.866_{-0.004}^{+0.003}$	&	$0.869_{-0.004}^{+0.003}$	&	$0.867_{-0.005}^{+0.004}$	&	$0.859_{-0.010}^{+0.008}$\\
         & $\tau$ && 						$2.4_{-0.1}^{+0.1}$ & $1.8_{-0.2}^{+0.2}$	&	$2.6_{-0.3}^{+0.3}$	&	$2.6_{-0.2}^{+0.2}$	&	$2.0_{-0.2}^{+0.2}$	&	$1.3_{-0.3}^{+0.3}$\\
         \multirow{2}{*}{Edge} &  E & keV &$0.957_{-0.005}^{+0.004}$	& $0.934_{-0.007}^{+0.006}$	&	$0.947_{-0.010}^{+0.009}$	&	$0.958_{-0.007}^{+0.007}$	&	$0.949_{-0.008}^{+0.008}$	&	$0.929_{-0.013}^{+0.012}$ \\
         & $\tau$ &&						$3.6_{-0.1}^{+0.2}$ & $3.9_{-0.5}^{+0.5}$	&	$3.3_{-0.3}^{+0.3}$	&	$4.0_{-0.3}^{+0.3}$	&	$3.4_{-0.3}^{+0.3}$	&	$2.2_{-0.4}^{+0.3}$ \\
         \multirow{2}{*}{Edge} & E & keV &	$1.139_{-0.009}^{+0.012}$ & $1.06_{-0.03}^{+0.03}$	&	$1.10_{-0.02}^{+0.02}$	&	$1.16_{-0.01}^{+0.01}$	&	$1.09_{-0.02}^{+0.02}$	&	$1.03_{-0.02}^{+0.02}$ \\
        & $\tau$ &&							$2.2_{-0.2}^{+0.2}$ & $1.6_{-0.5}^{+0.5}$	&	$1.9_{-0.2}^{+0.2}$	&	$2.2_{-0.2}^{+0.3}$	&	$1.9_{-0.3}^{+0.4}$	&	$2.2_{-0.4}^{+0.4}$ \\
         \multirow{2}{*}{Edge} & E & keV &$1.38_{-0.02}^{+0.02}$ 	& $1.21_{-0.02}^{+0.02}$	&	$1.33_{-0.02}^{+0.02}$	&	$1.46_{-0.03}^{+0.53}$	&	$1.28_{-0.02}^{+0.02}$	&	$1.21_{-0.02}^{+0.03}$ \\
        & $\tau$ &&							$2.0_{-0.3}^{+0.3}$ & $4.1_{-0.8}^{+1.0}$	&	$2.1_{-0.3}^{+0.4}$	&	$2.7_{-0.7}^{+1.4}$	&	$3.0_{-0.6}^{+0.8}$	&	$3.7_{-0.8}^{+1.2}$ \\
        & $\chi^2$/dof & & 221.2/82 & 88.5/44 & 113/52 &  95.1/50 & 66.4/47 &  79.0/44 \\
        \hline
        \multicolumn{9}{c}{Day 38} \\ \hline
          \multirow{2}{*}{Blackbody} & kT & keV & $0.115_{-0.002}^{+0.002}$ & $0.117_{-0.004}^{+0.004}$	&	$0.122_{-0.003}^{+0.003}$	&	$0.117_{-0.002}^{+0.003}$	&	$0.114_{-0.002}^{+0.003}$	&	$0.116_{-0.004}^{+0.005}$ \\
        & Norm&                                 & $0.104_{-0.002}^{+0.002}$ & $0.069_{-0.003}^{+0.003}$	&	$0.131_{-0.004}^{+0.004}$	&	$0.151_{-0.004}^{+0.004}$	&	$0.118_{-0.004}^{+0.004}$	&	$0.055_{-0.003}^{+0.003}$ \\
          \multirow{2}{*}{Edge}& E   & keV & 	$0.766_{-0.003}^{+0.003}$ & $0.747_{-0.005}^{+0.005}$	&	$0.750_{-0.005}^{+0.005}$	&	$0.768_{-0.004}^{+0.004}$	&	$0.771_{-0.005}^{+0.006}$	&	$0.76_{-0.01}^{+0.01}$ \\
         & $\tau$ && 							$0.91_{-0.04}^{+0.05}$ & $0.93_{-0.09}^{+0.09}$	&	$0.88_{-0.08}^{+0.08}$	&	$0.99_{-0.06}^{+0.06}$	&	$0.83_{-0.07}^{+0.07}$	&	$0.7_{-0.2}^{+0.1}$ \\ 	
          \multirow{2}{*}{Edge}& E & keV & 		$0.863_{-0.003}^{+0.003}$ & $0.85_{-0.01}^{+0.01}$	&	$0.840_{-0.006}^{+0.006}$	&	$0.869_{-0.004}^{+0.003}$	&	$0.868_{-0.006}^{+0.005}$	&	$0.84_{-0.02}^{+0.02}$ \\
         & $\tau$ &&							$2.2_{-0.2}^{+0.2}$ & $1.6_{-0.6}^{+0.6}$	&	$1.9_{-0.2}^{+0.2}$	&	$3.2_{-0.3}^{+0.3}$	&	$2.4_{-0.3}^{+0.3}$	&	$1.1_{-0.3}^{+0.4}$ \\ 
          \multirow{2}{*}{Edge}& E & keV &		$0.944_{-0.006}^{+0.007}$ & $0.92_{-0.02}^{+0.02}$	&	$0.922_{-0.009}^{+0.009}$	&	$0.96_{-0.01}^{+0.01}$	&	$0.95_{-0.01}^{+0.01}$	&	$0.92_{-0.02}^{+0.02}$ \\
         & $\tau$ &&							$3.2_{-0.2}^{+0.2}$ & $2.6_{-0.6}^{+0.5}$	&	$3.7_{-0.4}^{+0.4}$	&	$2.9_{-0.4}^{+0.4}$	&	$3.0_{-0.4}^{+0.4}$	&	$3.3_{-1.5}^{+0.4}$ \\
          \multirow{2}{*}{Edge}& E & keV &		$1.08_{-0.01}^{+0.01}$ & --	&	$1.06_{-0.02}^{+0.02}$	&	$1.14_{-0.02}^{+0.02}$	&	$1.11_{-0.02}^{+0.04}$	&	-- \\
         & $\tau$ && 							$2.1_{-0.2}^{+0.2}$ & --	&	$2.6_{-0.5}^{+0.5}$	&	$2.234_{-0.4}^{+0.4}$	&	$2.0_{-0.4}^{+1.0}$	&	-- \\
          \multirow{2}{*}{Edge}& E & keV &		$1.28_{-0.01}^{+0.01}$ & --	&	$1.26_{-0.02}^{+0.04}$	&	$1.41_{-0.02}^{+0.04}$	&	$1.29_{-0.04}^{+0.20}$	&	-- \\
         & $\tau$ && 							$4.1_{-0.6}^{+0.8}$ & --	&	$3.1_{-0.6}^{+0.8}$	&	$4.1_{-1.1}^{+5.8}$	&	$2.3_{-1.2}^{+0.9}$	&	-- \\
         & $\chi^2$/dof & &130.21/67 & 30.7/36 & 54.5/40 &  50.2/42 & 31.9/40 & 38.8/35\\ \hline
        \multicolumn{9}{c}{Day 54} \\ \hline
          \multirow{2}{*}{Blackbody}& kT & keV &  $0.107_{-0.001}^{+0.002}$  & $0.098_{-0.003}^{+0.003}$	&	$0.107_{-0.003}^{+0.003}$	&	$0.117_{-0.003}^{+0.003}$	&	$0.109_{-0.003}^{+0.003}$	&	$0.104_{-0.004}^{+0.004}$ \\
         & Norm&                            &     $0.064_{-0.002}^{+0.002}$  & $0.041_{-0.003}^{+0.003}$	&	$0.084_{-0.004}^{+0.004}$	&	$0.096_{-0.003}^{+0.003}$	&	$0.068_{-0.003}^{+0.003}$	&	$0.030_{-0.002}^{+0.002}$  \\
          \multirow{2}{*}{Edge}& E        & keV & 	$0.759_{-0.003}^{+0.002}$& $0.755_{-0.007}^{+0.007}$	&	$0.748_{-0.008}^{+0.007}$	&	$0.761_{-0.004}^{+0.004}$	&	$0.756_{-0.006}^{+0.006}$	&	$0.757_{-0.008}^{+0.006}$  \\
         & $\tau$ && 								$1.08_{-0.05}^{+0.05}$ & $1.0_{-0.1}^{+0.1}$	&	$0.9_{-0.1}^{+0.1}$	&	$1.21_{-0.08}^{+0.08}$	&	$0.9_{-0.1}^{+0.1}$	&	$0.9_{-0.1}^{+0.1}$  \\
          \multirow{2}{*}{Edge}& E & keV & 			$0.859_{-0.003}^{+0.003}$& $0.86_{-0.02}^{+0.02}$	&	$0.830_{-0.009}^{+0.009}$	&	$0.858_{-0.005}^{+0.004}$	&	$0.852_{-0.006}^{+0.006}$	&	$0.869_{-0.012}^{+0.007}$  \\
         & $\tau$ &&								$2.1_{-0.1}^{+0.1}$& $1.2_{-0.3}^{+0.3}$	&	$1.7_{-0.2}^{+0.2}$	&	$2.9_{-0.2}^{+0.2}$	&	$2.28_{-0.26}^{+0.26}$	&	$1.43_{-0.27}^{+0.24}$ \\
          \multirow{2}{*}{Edge}& E & keV &			$0.963_{-0.005}^{+0.004}$& $0.95_{-0.02}^{+0.02}$	&	$0.929_{-0.008}^{+0.008}$	&	$0.98_{-0.01}^{+0.01}$	&	$0.96_{-0.01}^{+0.01}$	&	$0.97_{-0.02}^{+0.01}$  \\
         & $\tau$ &&								$3.0_{-0.1}^{+0.2}$& $2.9_{-0.6}^{+0.5}$	&	$3.7_{-0.3}^{+0.3}$	&	$3.3_{-0.3}^{+0.3}$	&	$2.8_{-0.3}^{+0.3}$	&	$2.1_{-0.4}^{+1.0}$ \\
          \multirow{2}{*}{Edge}& E & keV &			$1.20_{-0.01}^{+0.01}$& $1.25_{-0.12}^{+0.07}$	&	$1.17_{-0.02}^{+0.02}$	&	$1.19_{-0.02}^{+0.02}$	&	$1.18_{-0.02}^{+0.02}$	&	$1.12_{-0.02}^{+0.17}$  \\
         & $\tau$ &&								$2.5_{-0.2}^{+0.3}$& $1.1_{-0.6}^{+1.3}$	&	$2.8_{-0.5}^{+0.7}$	&	$3.2_{-0.5}^{+0.6}$	&	$2.6_{-0.5}^{+0.6}$	&	$1.4_{-0.6}^{+0.6}$ \\
         & $\chi^2$/dof && 145.35/71 & 30.6/37 & 74.6/40 & 74.9/41 & 38.96/40 & 43.1/37\\
         \hline
    \end{tabular}
   
\end{table*}

\subsection{UV spectral analysis}
\begin{figure}
    \centering
    \includegraphics[width=\columnwidth]{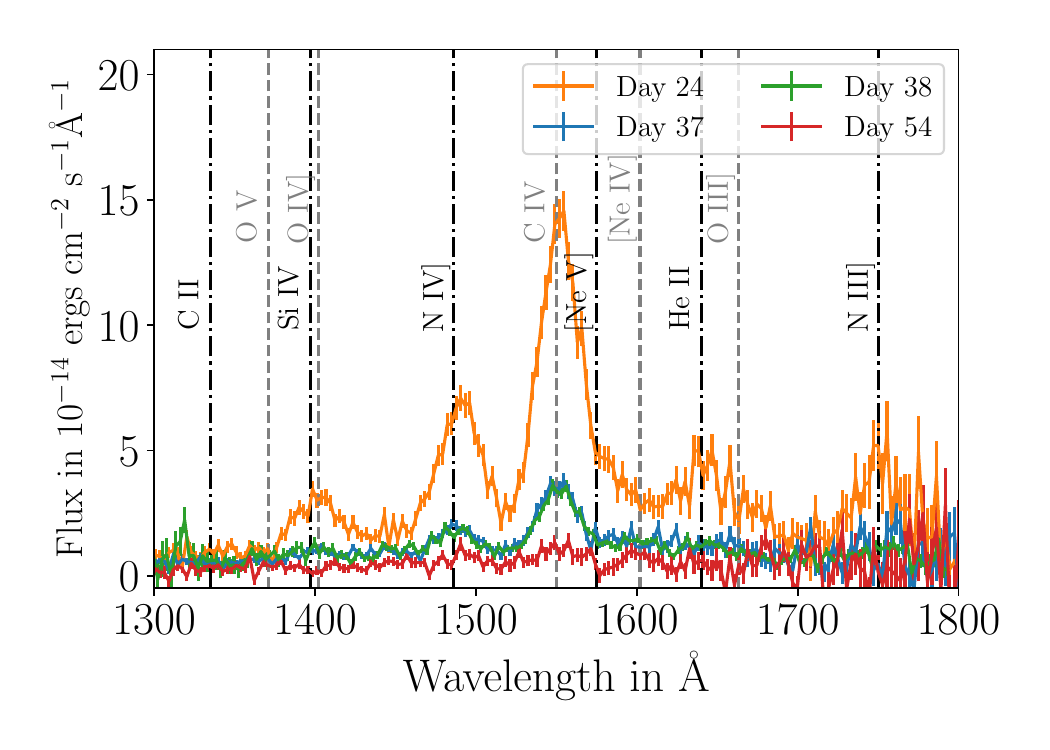}
    \caption{UV spectra for different epochs of observations. The identified emission lines (also noted in Table~\ref{tab:uv_spec_pars}) are marked as vertical dashed/dot-dash lines to guide the eye.  }\label{fig:uv_spec_lines}
\end{figure}

The UV spectra for all the epochs show clear emission features. We model these features with narrow Gaussian functions and include a blackbody component to model the continuum. 
 The full width at half maximum (FWHM) is estimated from the brightest features, and assumed to be the same for all the other features.  The typical observed width  is close to the grating resolution.  We identify the observed narrow features in the spectra with emission lines from various elements as suggested by \citet{vandingham1997MNRAS.290...87V}.  The UV spectra for each epoch are shown in Figure~\ref{fig:uv_spec_lines} with identified emission features indicated as vertical lines. The line flux of these features are noted in Table~\ref{tab:uv_spec_pars}. The strongest emission features (\ion{N}{IV}] and \ion{C}{IV}) are seen in all observations, while some of the faint features are only seen at brighter epoch or in observations with both gratings. As the source  UV flux decreases across  different observations, we find that the line fluxes are also decreasing.

\begin{table*}
    \centering
    \caption{Fluxes of the emission features seen in the UV spectra at various epochs.  }\label{tab:uv_spec_pars}
    \begin{tabular}{lcrrrr}
        \hline
       Spectral line & Observed wavelength (\AA)  & \multicolumn{4}{c}{Flux in $10^{-13}$ erg cm$^{-2}$ s$^{-1}$}  \\
         & & Day 24 & Day 37 & Day 38 & Day 54\\ \hline
         \ion{C}{II} (1335) & 1325 & -- & 1.40 & --&-- \\ \hline
         \ion{O}{V} (1371) & 1363 & -- & 0.90 &-- &-- \\ \hline
       \ion{Si}{IV}/\ion{O}{IV}] (1402) & 1389	&	13.4 & 2.98 &3.03 &--\\ \hline
        Unidentified & 1448 & 3.10 & 2.13 & 1.45 & 0.98 \\ \hline
       \ion{N}{IV}]     (1486) & 1490	& 20.9 & 6.73 &5.69 & 2.04 \\ \hline
       \ion{C}{IV}  (1549) &  1551   & 42.4 & 12.2 & 10.4 & 3.72  \\ \hline
       {[\ion{Ne}{V}]} (1575) & \multirow{2}{*}{1588} &  \multirow{2}{*}{11.8}& \multirow{2}{*}{3.44} & \multirow{2}{*}{3.23} & \multirow{2}{*}{1.81} \\
       {[\ion{Ne}{IV}]} (1602) \\ \hline
       \ion{He}{II}  (1640) & 1645 & 13.9 & 5.02 & 3.85 &-- \\ \hline
       \ion{O}{III}]  (1665) & 1670 & 7.14 &--&--&--\\ \hline
       \ion{N}{III}]  (1750) & 1753 & 11.4 & 4.42 & 1.83 &-- \\ \hline
    \end{tabular}
    \begin{flushleft}
    Note: Typical uncertainty in the flux estimates is 20--30\%
    \end{flushleft}
\end{table*}

\section{Results and Discussion}\label{sec:results}

We observed the nova \src\ during the supersoft phase with \astr\ in X-ray and UV wavelengths and have conducted a detailed spectral and timing study of the source. 
\subsection{Pulsations in the source}

The X-ray observations of the source have shown a strong pulse profile after the peak of the SSS phase and which is also visible even 10 months after the initial eruption \citep{Page2022ATel15317....1P}. We estimate that the pulse period of the source is 501.44--501.53~s (Table~\ref{tab:period_x-ray}) which is statistically different from the estimates using \chan\ \citep{Drake2021ApJ...922L..42D} and \nic\ \citep{Orio2022ApJ...932...45O}. Since there are considerable simultaneous observations of the source, we sampled the \chan\ observation at SXT observation windows and computed the LS periodogram which matches the SXT period. And shifting the sampling window resulted in an intermediate period indicating that for this source, the periodogram is sensitive to the sampling window as the size of the sampling window is similar to the period. 
Additionally, the individual pulses are quite different causing a strong dependence on which pulse is sampled and at what phase.
 Thus we note that the uncertainty in the period may be dominated by the sampling of the source activity and the typical uncertainty estimation techniques \citep[used by ][and outlined in Appendix~\ref{app:error}]{Drake2021ApJ...922L..42D} fail to incorporate the intrinsic variability of the individual pulses and the deviation of the pulse phase from a sinusoid function. The difference in the observed period is an artifact of the subset of the pulses being observed and thus a continuous observation (e.g. \chan\ observation) is paramount to estimate the \emph{true} period of the source.

The UV observations do not show  pulsations at the period seen in X-rays (see Figure~\ref{fig:uvit_lsp}). To verify if this is due to lack of UV sensitivity or due to inherent lack of pulsations, we simulated UV lightcurves and injected pulsations at various strengths (relative to the mean count rate). We find that our lightcurves are sensitive to detect a pulse fraction of 0.05 for a conservative false alarm probability of 1 \%. Thus we conclude that UV pulsations are weaker than 0.05 (the X-ray pulsations are roughly 0.5--0.6 at a similar epoch).  The optical observations in \citet{patterson2022ApJ...940L..56P} indicate that the pulse fraction for the 501.4~s modulation evolves from  0.005 on day 12 to 0.04 on day 350 (0.01 mag at V=12 to 0.09 mag at V=17 converted to the peak to trough pulse fraction).  

 The X-ray pulsations are observed strongly after the peak of the SSS phase \citep[both in our analysis and][]{Drake2021ApJ...922L..42D} while the optical modulations are observed as early as day 12 of the eruption \citep{patterson2022ApJ...940L..56P}. The continuation of the X-ray modulations in the quiescence period strongly indicates that they are due to the hotter regions on the polar caps and that the observed periodicity is likely due to the rotation period of an intermediate polar.  However, the modulation observed in outburst is of such amplitude and at such high luminosity that it cannot be solely due to accretion.   
While the reason for the non-detection of the pulsations in the UV is not very clear, one possibility could be a significant smearing of the modulation due to flickering caused by a re-accretion process.   Based on the X-ray and UV timing analysis, we can conclude that the WD atmosphere (the typical source of the X-ray emission) is inhomogenous and is modulated with the rotation period of the WD while the UV emission is not significantly modulated at the same time, perhaps, indicating a distinct origin of the UV emission.

\subsection{Evolution of the spectrum}

The source emission in X-rays can be primarily described as a blackbody component with multiple absorption edges. We detect a relatively hotter blackbody \citep[as compared to ][]{Drake2021ApJ...922L..42D, Orio2022ApJ...932...45O}. Across the three epochs after the peak of the SSS phase, the temperature is found to be roughly constant. The normalisation of the blackbody (indicative of the size of the emission region) follows the trend suggested by the SSS evolution. 

The UV spectra of the source indicate an evolution distinct from the X-ray evolution. Throughout the SSS phase, the UV flux is observed to decrease, which is also seen in the evolution of individual emission features. Some of the fainter features are not observed in later observations possibly due to the lack of statistics. The decrease in the UV distinct from the development of the soft X-ray flux indicates the UV fluxes originate in a region different from the soft X-rays, most likely the inner regions of the nova ejecta.

\subsection{Evolution of optical depth as a function of the pulse phase}

 Noting a clear evolution of the X-ray spectrum as a function of the pulse phase (see Figure~\ref{fig:ratio_spec} for a model independent depiction of the same), we conducted a pulse phase resolved spectral study to investigate the dependence of the spectral parameters as a function of the pulse phase. We find that the pulse variation of the flux can be characterised as the variation in the normalisation of the blackbody component (and thus the area of the emitting region).  Our spectral modelling leads to the incorporation of a few X-ray absorption edges to bring the $\chi^2$ closer to acceptable values. Additionally, we find a marginal evidence of variation of the optical depth of the observed edges as a function of the pulse phase.  The presence of absorption edges in the low resolution SXT spectra may, however, be compromised by the complexity of the model with large number of components used here and would require higher resolution X-ray spectra for confirmation in future observations of such novae.

\section{Conclusions}
Our analysis of nearly-simultaneous soft X-ray and FUV observations of V1674~Her confirms the presence of a 501.4--501.5~s period in the X-rays at the peak of the SSS phase. However, no periodicity is detected in the FUV. The shape of the phase-folded X-ray light curves varies from one epoch to another across the three epochs observed during the peak of the SSS phase. We also report the presence of absorption features in the X-rays, and strong line emission features identified with Si, N, and O in the FUV. The non-detection of the $\sim$501~s modulation in UV wavelengths and the disjoint evolution of FUV and X-ray fluxes during the period of our observations, indicate the FUV and X-ray emissions most likely arise from different regions.

\section*{Acknowledgements}

We thank the Indian Space Research Organisation for scheduling the observations and the Indian Space Science Data Centre (ISSDC) for making the data available.  
This work has been performed utilizing the calibration data-bases and auxillary analysis tools developed, maintained and distributed by \astr-SXT team with members from various institutions in India and abroad and the  SXT Payload Operation Center (POC) at the TIFR, Mumbai for the pipeline reduction. This publication uses UVIT data processed by the payload operations center at IIA. The UVIT is built in collaboration between IIA, IUCAA, TIFR, ISRO, and CSA.
The work has also made use of software, and/or web tools obtained from NASA's High Energy Astrophysics Science Archive Research Center (HEASARC), a service of the Goddard Space Flight Center and the Smithsonian Astrophysical Observatory. Kulinder Pal Singh thanks the Indian National Science Academy for support under the INSA Senior Scientist Programme. GJML is a member of CIC-CONICET (Argentina) and acknowledge support from grant ANPCYT-PICT 0901/2017. JJD was funded by NASA contract NAS8-03060 to the \chan\ X-ray Center and thanks the Director, Pat Slane, for continuing advice and support. This work made use of \textsc{astropy}:\footnote{http://www.astropy.org} a community-developed core Python package and an ecosystem of tools and resources for astronomy \citep{2013A&A...558A..33A,2018AJ....156..123A,2022ApJ...935..167A}. 

\section*{Facilities}

\astr, \nic,  \chan.

\section*{Data Availability}

\astr\ observations are available at \url{https://astrobrowse.issdc.gov.in/astro_archive/archive/Home.jsp}. The observations can be searched for using the observation IDs mentioned in Table~\ref{tab:obs_log_astr}.



\bibliographystyle{mnras}
\bibliography{ref} 




\appendix

\section{Estimating the uncertainty in the period from X-ray observations}\label{app:error}

To determine the uncertainty in the  estimated period we followed the procedure outlined here. 

\begin{enumerate}
   
    \item The given lightcurve is fitted with a sinusoid function to determine the best-fit period. 
    \item Using the best-fit sinusoid parameters, we simulate 5000 lightcurves with exposure and total duration similar to the original lightcurve. The individual time stamps for the simulated lightcurve need not correspond to the time stamps of the original lightcurve. The count rate for each time stamp is determined using a sinusoid function as determined in step (i) 
    \item Each flux point was randomised by including an uncertainty sampled from a Gaussian distribution. 
    \item For each simulated lightcurve, we computed the LS periodogram and determined the position of the peak. The standard deviation of the  periods is assumed to be the uncertainty in the period estimate. 
\end{enumerate}

\begin{figure}
    \centering
    \includegraphics[width=\columnwidth]{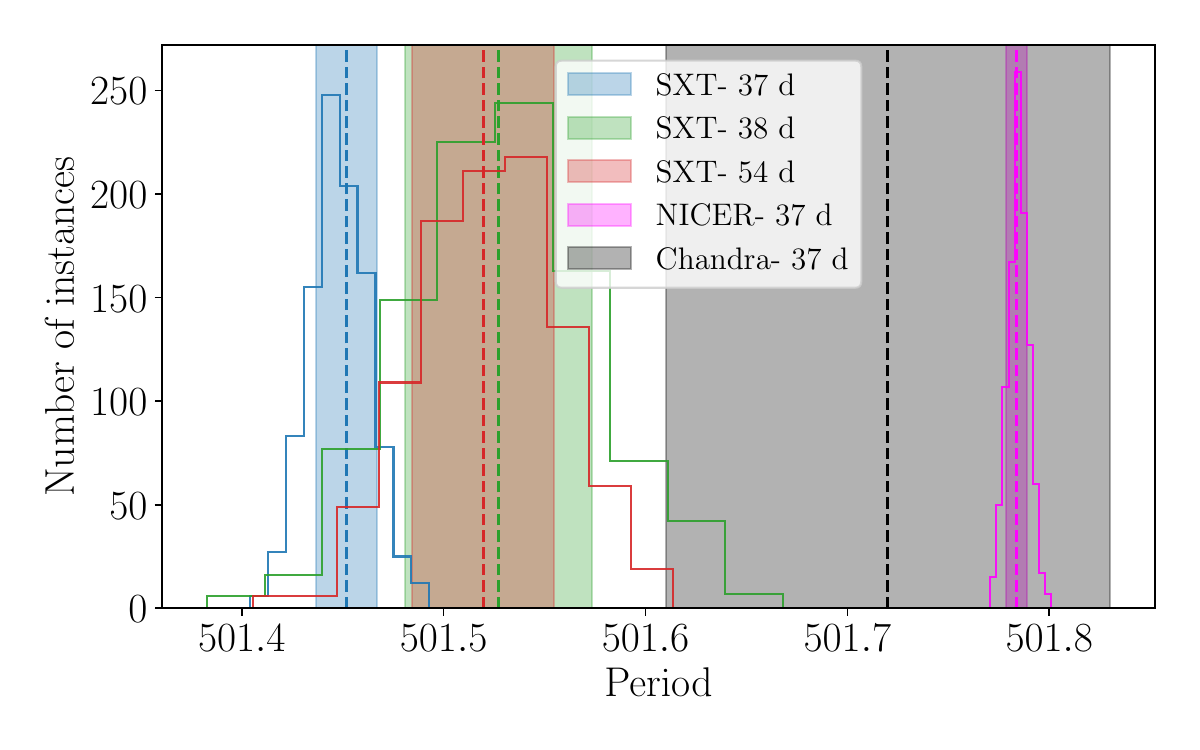}
    \caption{Distribution of the period estimates from the simulated lightcurves. Vertical dashed lines indicate the period derived from the LS periodogram of the observations and the shaded region  represents the standard deviation of the distributions. }\label{fig:hist_xray_simuls}
\end{figure}

We determined the distribution for all the X-ray observations which showed significant periodicity (namely SXT-Day 37, SXT-Day 38 and SXT-Day 54 and NICER). The standard deviation of the distribution corresponds to 68\% uncertainty estimate. The distributions for different observations is shown in Figure~\ref{fig:hist_xray_simuls}.
Another important metric to estimate the lower limit on the uncertainty could be the Rayleigh criterion. The criterion encapsulates a typical error introduced if one period is missed/extra in counting the number of periods between two epochs. In the present case, the maximum separation between two epochs with detected period is $54-37=17$ days. Corresponding to a rough period of 501.7 s, we have 2927 cycles between the two epochs. And thus the expected error from the Rayleigh's criterion is $\approx 501.7\times1/2927 = 0.17$~s. This error doesn't incorporate the signal to noise ratio and therefore may be a overestimation of the lower limit \citep{2003ASPC..292..383S}. 

\bsp\	
\label{lastpage}
\end{document}